# Towards a simple, comprehensive model of regular earthquakes and slow slip events, part II: two-dimensional model

Naum I. Gershenzon* and Thomas Skinner

*Corresponding author
E-mail address: naum.gershenzon@wright.edu
Physics Department, Wright State University, 3640 Colonel Glenn Highway Dayton, OH 45435

**Abstract**

Although our existing one-dimensional (1D) model provides a successful quantitative description of rupture events, a 1D description is somewhat limited. We therefore derive a two-dimensional (2D) model which allows us to investigate characteristics of earthquakes (EQs) and slow slip events (SSEs) that are only apparent in a second dimension. We find that the leading edge of an EQ rupture in the direction of the global shear stress (*x*-direction) is wider in the plane of the crustal fault (*y*-direction) than the trailing edge. The direction of the slip velocity is primarily in the *x*-direction. EQ ruptures expand in both the *x*- and *y*-directions. In SSEs, the rupture also expands in both directions for a short period of time, then, after pulses are formed, there is no further expandsion, i.e., the pulse shape remains practically unchanged. The 2D simulations show the seismic moment (*M*) versus time (*T*) scaling law may be expressed by the relation $M \propto T^\lambda$, where $1 < \lambda < 3$. For EQs, $\lambda$ is approximately equal to 3, while for SSEs, $\lambda$ tends to 1 as the pulses develop and become fully established. The 2D model quantitatively describes the parameters of a seismic pulse in episodic tremor and slip (ETS) phenomena and also explains important features of non-volcanic tremor associated with this pulse, such as reverse tremor migration. The 2D model confirms the basic findings obtained previously by the 1D model: (i) The type of a seismic event, EQ or SSE, and the type of a rupture, crack-like or pulse-like, are determined solely by the fault strength, the ratio of the shear to normal stress, and the gradient in this ratio; (ii) The shape of a rupture and position of the maximum slip is determined by the spatial distribution of the initial stress.

**INTRODUCTION**

Part I of this article (Gershenzon et al, 2019) was devoted to developing a one-dimensional (1D) model describing the major characteristics of a rupture, ranging from regular earthquakes (EQs) to slow slip events (SSEs). We showed that the type of a seismic event, that is, regular EQ or SSE, is determined by the shear to normal stress ratio, its gradient, and by the fault strength. The key determinant for the occurrence of a regular EQ rather than a SSE was found to be the abrupt decrease of friction at slip velocities $\geq 0.1$ m/s due to flash heating (Yuan and Prakash, 2008; Beeler et al, 2008; Goldsby and Tullis, 2011) or other mechanisms (Brodsky and Kanamori, 2001). If the stress ratio and gradient in this ratio are large enough to accelerate the slip velocity to ~0.1 m/s then a regular EQ occurs. For lower values, all types of SSEs are possible. We also showed that the rupture velocity for regular EQs ranges from a fraction of the shear wave velocity ($c_s$) up to the supershear velocity. The range for SSEs is even wider, varying from $10^{-7} c_s$ to $10^{-1} c_s$. The maximum slip velocity for regular EQs is $\geq 1$ m/s. For SSEs, slip velocities range from a few cm/year up to 0.1 m/s. Stress drop varies in the range 1% to 10% of the initial shear stress for regular EQs and between 0.1% and 10% for SSEs. Generally, stress drop in SSEs is one to three orders of magnitude less than in EQs. We also determined the conditions that distinguish between the occurrence of the two main rupture modes, i.e., crack-like and self-healing pulse modes. The type of rupture is determined by the stress ratio and by details of the stress heterogeneity. The rupture is crack-like for large enough values of the ratio and its gradient and pulse-like for smaller values. Heterogeneity in the stress spatial distribution controls the position of maximum slip.

The advantages of our 1D model are 1) for some initial conditions and boundary values the system of equations describing the model admits analytical solutions, which is important for establishing basic characteristics of seismic events and for testing the results of computer simulations; 2) 1D simulations place minimal demands on computer memory and are not computationally intensive; 3) typically, the basic rupture characteristics are more pronounced and evident in 1D simulations than in 2D or 3D. However, some characteristics of the rupture could not be accounted for in the 1D model.

Here, in the Part II, we develop a two-dimensional (2D) model of a rupture in the plane along the crustal fault. The 2D model follows the basic format of the 1D model in that it is also derived from the Frenkel-Kontorova (FK) model (Kontorova and Frenkel, 1938) and incorporates the rate-state (Dieterich, 1979; Ruina, 1983) and flash heating (Yuan and Prakash, 2008; Beeler et al, 2008; Goldsby and Tullis, 2011) frictional laws. The model is consistent with results of the previous 1D model and is suitable for describing the kinematic and dynamic parameters of a rupture, i.e., the spatial and temporal redistribution of shear to normal stress ratio, slip velocity, slip, and rupture velocity during EQs and SSEs. Besides deriving the system of equations for our 2D description, this article is focused on additional general features of seismic ruptures that emerge from the model. We can now account for important features of episodic tremor and slip (ETS) phenomena, such as reverse tremor migration (Houston et al, 2011), and are able to consider the seismic moment ($M_0$) versus rupture time ($T$) scaling law for EQs and SSEs. The conventional point of view of a scaling law based on the empirical data is that $M \propto T^3$ for regular EQs and $M \propto T$ for the SSEs (Ide et al., 2007, 2008; Ide, 2014). Several other models have also been developed explaining these scaling relations (Ide, 2008; Ben-Zion, 2012; Colella et al., 2013; Liu, 2014; Gomberg et al, 2016).

Section Model describes the model and the system of governing equations. Section Results and Discussion presents the results of simulations, which are discussed in the context of known conventional results. The final section summarizes our findings.

**MODEL**

Complete details of the 1D model can be found in our previous publications (Gershenzon, 1994; Gershenzon et al, 2009, 2015, 2019; Gershenzon and Bambakidis, 2012). Part I also includes a synopsis of its main features. The 2D model requires modifications of the governing equations, as follows. To begin with, the sine-Gordon (sG) equation we obtained as Eq. (4) in Part I, which describes the relevant processes at the microscopic level, now has to include a term in the *y*-direction, giving

$$\frac{\partial^2 \tilde{u}}{\partial \tilde{t}^2} - \frac{\partial^2 \tilde{u}}{\partial \tilde{x}^2} + \frac{\partial^2 \tilde{u}}{\partial \tilde{y}^2} + \sin \tilde{u} = \tilde{\sigma}_s - \tilde{\sigma}_f \qquad (1)$$

where (see Part I) $\tilde{u}$ is the shift of an asperity relative to its equilibrium position in the slip direction in units of $b_0 / 2\pi$ ($b_0$ is the typical distance between asperities), $\tilde{x}$ and $\tilde{y}$ are coordinates in the fault plane in units of $[\frac{2G}{2\pi \Sigma_N (1-\nu)}]^{1/2} b_0$ ($G$ is the shear modulus, $\nu$ is the Poisson ratio, and $\Sigma_N$ is the normal stress), $\tilde{t}$ is time in units of $[\frac{2G}{2\pi \Sigma_N (1-\nu)}]^{1/2} \frac{b_0}{c}$, $\tilde{\sigma}_s$ is shear stress in units of $\Sigma_N$, $\tilde{\sigma}_f = \mu \Sigma_N$ ($\mu$ is friction force per unit area), $c^2 = \frac{2G}{\rho(1-\nu)} \equiv \frac{c_l^2(1-2\nu)}{(1-\nu)^2}$, where $\rho$ is the volume density, and $c_l$ is the longitudinal acoustic velocity, with $c_s < c < c_l$, where $c_s$ is the shear wave velocity.

### *The sine-Gordon modulation equation*

As discussed in Part 1, friction at the macroscopic level is more appropriately described by what are known as modulation equations, built upon the equations governing the underlying microscopic processes. Therefore, starting with the sG equation, we employ Whitham's method (Whitham, 1974) to derive the relevant system of modulation equations.

The method requires us to consider first the modulation equations for the homogeneous sG equation (right hand side equal to zero). We consider a solution of Eq. (1) in the form of a wave packet $u = \Psi(\theta)$, where $\theta = k_x x + k_y y - \omega t$, $k_x$ and $k_x$ are the wave numbers in the $x$ and $y$ directions, and $\omega$ is the frequency. Then the lagrangian, $L$, of Eq. (1) has the form:

$$L = \frac{1}{2}(\omega^2 - k^2)^{1/2} (\frac{d\Psi}{d\theta})^2 - (1 - \cos \Psi),$$

where $k^2 = k_x^2 + k_y^2$.

The lagrangian averaged over an oscillation period is

$$\bar{L} = \frac{1}{2\pi}[2(\omega^2 - k^2)]^{1/2} \int [A - (1 - \cos \Psi)]^{1/2} d\Psi - A,$$

where $A$ is an integration constant. In accordance with modulation theory (Whitham, 1974), the averaged lagrangian should satisfy the following system of equations, where, in what follows, the notation $\partial_i$ denotes the partial derivative with respect to the associated variable $i$:

$$\partial_A \overline{L} = 0 \tag{2}$$

$$(\partial_t \partial_\omega - \partial_x \partial_{k_x} - \partial_y \partial_{k_y}) \overline{L} = 0 \tag{3}$$

$$\partial_t k_x + \partial_x \omega = 0, \tag{4}$$

$$\partial_t k_y + \partial_y \omega = 0, \tag{5}$$

$$\partial_y k_x + \partial_x k_y = 0 \tag{6}$$

From Eq. (2) we find:

$$k(U, m) = \frac{\pi}{K(m)[m(U^2 - 1)]^{1/2}}, \tag{7}$$

where $K(m)$ is the complete elliptic integral of the first kind, $m$ is the modulus of the Jacoby elliptic function ($0 < m \leq 1$), and $U = \omega/k$. Calculating the derivatives in Equations (2) – (6), rearranging terms, and using Eq.(7) allows us to reduce the system of equations to the following three equations in terms of the independent variables $m$, $U$, and $\alpha$ ($\alpha$ is the angle between $k_x$ and $k$):

$$(\frac{\eta}{U^2-1}\partial_t + \frac{\eta U}{U^2-1}\cos\alpha\, \partial_x + \frac{\eta U}{U^2-1}\sin\alpha\, \partial_y)U + (\frac{U}{2m}\partial_t + \frac{1}{2m}\cos\alpha\, \partial_x + \frac{m_y}{2m}\sin\alpha\, \partial_y)m = 0, \tag{8}$$

$$(\frac{U}{U^2-1}\partial_t + \frac{1}{U^2-1}\cos\alpha\, \partial_x)U + (\frac{\eta}{2mm_1}\partial_t + \frac{\eta U}{2mm_1}\cos\alpha\, \partial_x)m = 0, \tag{9}$$

$$(\frac{U}{U^2-1}\partial_t + \frac{1}{U^2-1}\sin\alpha\, \partial_y)U + (\frac{\eta}{2mm_1}\partial_t + \frac{\eta U}{2mm_1}\sin\alpha\, \partial_y)m = 0, \tag{10}$$

where $m_1 = 1 - m$, $\eta = E/K$, $E(m)$ is the complete elliptic integrals of the second kind, and $U$ is the wave velocity in units of velocity $c$ ($-1 \leq U \leq 1$).

Whitham also showed how to derive modulation equations for the inhomogeneous sG equation (Whitham, 1974). Following his procedure results in the following modification to Eq. (8), while Eqs. (9) and (10) remain the same:

$$(\frac{\eta}{U^2-1}\partial_t + \frac{\eta U}{U^2-1}\cos\alpha\,\partial_x + \frac{\eta U}{U^2-1}\sin\alpha\,\partial_y)U + (\frac{U}{2m}\partial_t + \frac{1}{2m}\cos\alpha\,\partial_x + \frac{m_y}{2m}\sin\alpha\,\partial_y)m =$$
$$-\frac{\pi}{4}\frac{[m(U^2-1)]^{1/2}}{K}(\Sigma_s - \mu - \Sigma_{srd}), \quad (11)$$

where the macroscopic shear stress $\Sigma_s = \frac{1}{2\pi}\int_0^{2\pi}\sigma_s\,d\xi$., $\Sigma_{srd} = \frac{G}{2c_s}\frac{\partial u}{\partial t}$ is the term which accounts for seismic radiation damping (Rice, 1993). It could be considered as an addition to the effective friction. In dimensionless units

$$\Sigma_{srd} = (\frac{G}{8\pi\Sigma_N})^{1/2}\frac{\partial u}{\partial t}. \quad (12)$$

As in Part I, the variables $m(t,x,y)$ and $U(t,x,y)$ are associated with the macroscopic variables $\Sigma_S$ (in units of $\Sigma_N$) and slip velocity $W$ (in units of $c(\frac{\Sigma_N}{2\pi G}\frac{1-\nu}{2})^{1/2}$) by the following relations:

$$\Sigma_S = \frac{\pi}{K[m(1-U^2)]^{1/2}},$$
$$W = U\Sigma_S. \quad (13)$$

Thus, the system of equations (9) – (13) describes the spatiotemporal redistribution of shear stress and slip velocity in the 2D model.

*Friction*

Although friction is incorporated in the present model in the same way as for the 1D model, we repeat its description for convenience. For slip velocity $W \leq 0.1$ m/s, we will use the well-known Dieterich-Ruina rate-state model (Dieterich, 1979; Ruina, 1983) for the coefficient of friction, $\mu$:

$$\mu(W,\theta) = \mu_0 + a\ln(\frac{W}{W_0}) + b\ln(\frac{W_0\theta}{D_c})$$
$$\frac{d\theta}{dt} = 1 - \frac{W\theta}{D_c}, \quad (14)$$

where $a$ and $b$ are empirical dimensionless constants, $\theta$ is the state variable in units of $b_0 G/(2\pi\Sigma_N c)$, and $D_c$ is the characteristic slip distance. We assume that $b_0/(2\pi D_c) \approx 1$. When $\dot\theta = 0$, we have the steady state values $\theta_{ss} = D_c/W$, in which case, the steady state friction $\mu_{ss}$ is

$$\mu(W, \theta_{ss}) \equiv \mu_{ss}(W) = \mu_0 + (a-b)\ln(\frac{W}{W_0}),$$

giving $\mu_{ss}(W_0) = \mu_0$. Representative reference values for $\mu_0$ and $W_0$ are provided in Table 1.

Table 1. Values of the parameters used in simulations

| variable | value | variable | value |
|---|---|---|---|
| $G$ | 30 GPa | $\mu_{flash}$ | 0.2 |
| $\nu$ | 0.3 | $W_{flash}$ | 0.1 m/s |
| $c_l$ | 6 km/s | $W_0$ | 3 cm/year |
| $c_s$ | 3.67 km/s | $\Sigma_N$ | 30 MPa |
| $c$ | 5.66 km/s | $\Sigma_{S0}$ | 0.15÷0.9 |
| $b_0$ | 3 cm | $\delta\Sigma$ | 0.001÷0.1 |
| $a$ | 0.01 | $\beta$ | 0.1 |
| $b$ | 0.015 | $\zeta$ | 0.01 |

For $W > 0.1$ m/s, friction abruptly decreases as a function of $W$, due to flash heating and then sharply increases with further increase of $W$ due to radiation damping effects from the resulting seismic waves (see Part I), according to

$$\mu(W) = [(\mu_{ss}(W_{flash}) - \mu_{flash})\frac{W_{flash}}{W} + \mu_{flash} + (\frac{G}{8\pi\Sigma_N})^{1/2}W, \tag{15}$$

An example of the dependence of friction on slip velocity, $\mu(W)$, is shown in Fig. 1 for the particular value $\theta = \theta_{ss}$ (steady state).

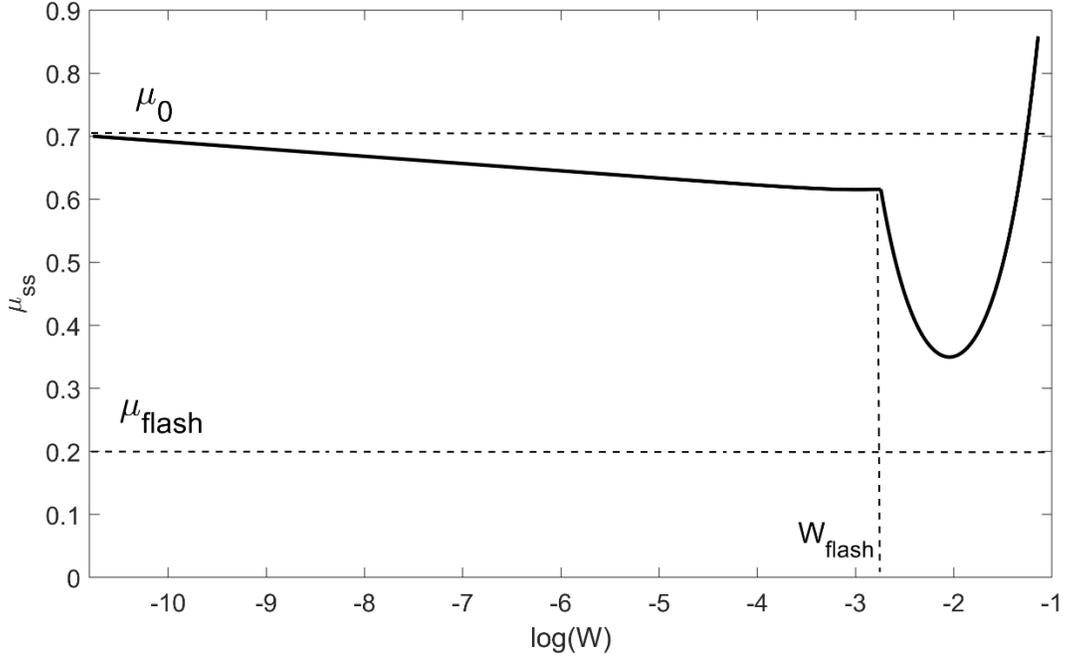

Figure 1. The steady state $\theta = \theta_{ss}$ is used as an example of the velocity-dependent friction coefficient, plotted as a function of (dimensionless) slip velocity, $W$: $\mu(W, \theta = D_c/W) \equiv \mu_{ss}(W)$ as given in Eq. (10) for $W \leq W_{flash}$ and $\mu_{ss}(W) \equiv \mu(W)$ from Eq. (11) for $W > W_{flash}$, using parameters listed in Table 1 which give a velocity-weakening ($a < b$) example for rate-state friction. The value $\mu_0 = \mu_{SS}(W_0)$, where $W_0 = 2.4 \times 10^{-11}$ in dimensionless units has been chosen as a representative value for the average slip velocity along a fault, corresponding to ~3 cm/yr.

*Initial conditions used in the model*

To solve the system of Eqs. (9) – (15), initial conditions must be specified. The initial slip velocity $W_0 = W(x, t = 0)$ is chosen to be the average slip velocity along a fault. A typical value is $W_0 = 3$ cm/yr. In contrast to conventional approaches, we will use spatially inhomogeneous shear stress as an initial condition, since sliding instabilities develop in locations with a gradient in the ratio of shear to effective normal stress (Gershenzon, 2019).

Two configurations of the initial stress distribution will be used for the simulations (shear stress applied, without loss of generality, in the positive *x*-direction):

$$\Sigma_s(x, t = 0) = \Sigma_{s0} - \frac{\delta \Sigma_a}{\pi} \tan^{-1}(\beta x) e^{-\zeta_x |x| - \zeta_{ya} |y|} \qquad (16a)$$

$$\Sigma_s(x, t = 0) = \Sigma_{s0} + \frac{\delta \Sigma_b}{2 \cosh(\zeta_x |x| + \zeta_{yb} |y|\}]}, \qquad (16b)$$

where $\Sigma_{S0}$ is the value of the average shear stress in the distribution, $\delta\Sigma_a$ and $\delta\Sigma_b$ are the constants defining the stress gradient, $\beta$ regulates the slope of the profile, and $\zeta_x$, $\zeta_{ya}$ and $\zeta_{yb}$ determine the spatial extent of variations in the stress distribution in the *x*- and *y*-direction, respectively. The first configuration mimics the spatial distribution of shear stress around an obstacle, which prevents relative sliding of the plates along a fault. Stress accumulates on one side of the obstacle and there is a stress deficit on the other side. The second configuration approximates effects due to localized reductions in the effective normal stress, due, for example, to changes in pore pressure. In order to properly compare the simulation results, the difference between maximum and minimum stress values in both cases is set equal, which requires $\delta\Sigma_b \equiv \delta\Sigma = \dfrac{\pi}{2} \dfrac{e^{\zeta_x |x_{max}|}}{\tan^{-1}(\beta |x_{max}|)} \delta\Sigma_a$ in Eq. (16b), where $x_{max}$ is the location of the maximum in the stress profile of Eq. (16a); $\delta\Sigma$ is a associated constant defined in Part I. To make the spatial extent in the *y*-directions approximately equal for both profiles, we require $\zeta_{yb} = \zeta_{ya}\,\text{arccosh}(e)$.

The required initial conditions for the variables $m(x,y,t=0)$ and $U(x,y,t=0)$ of Eqs. (9–11) are then obtained from the initial stress profiles given in Eqs.(16) together with the defining relations for shear stress and slip velocity of Eqs. (12–13). Values of the parameters used in simulations are given in Table 1. A FORTRAN code was developed to obtain numerical solutions. The code allows one to study the evolution of frictional processes employing the predictor-corrector scheme.

## RESULTS AND DISCUSSION

In what follows, we assume the rupture is initiated at $t=0$ at the point $x=0$, $y=0$ where the gradient in the stress ratio is a maximum.

### *Regular EQs*

As shown in Part I, a regular EQ appears if the shear stress and stress ratio are large enough to generate a slip with velocity $\geq 0.1$ m/s. Then flash heating abruptly reduces friction, triggering an increase in the slip velocity to values $\geq 1$ m/s. Further increase is bounded by the abrupt increase of friction due to the appearance of seismic waves.

Consider the spatially localized inhomogeneity in the initial stress given by Eq. 16a. Simulations for values $\Sigma_{0S} \equiv \mu_0 = 0.7$, $\delta\Sigma_a = 0.05 \cdot 2.3123 = 0.1156$ (equivalent to $\delta\Sigma = 0.05$), $\zeta_x = 0.025$, $\zeta_{ya} = 0.05$, and $\beta = 0.1$ are plotted in Fig. 2 at $t = 4.1$ s. The initial stress profile is shown in Fig. 2a. The results are typical of a regular EQ. The rupture is crack-like and propagates in both the $x$- and $y$-directions. The rupture velocity in the $x$-direction (direction of global tangential stress) is ~3.2 km/s and a few times larger than the rupture velocity in the $y$-direction. The region with reduced stress propagates in the direction of the initials stress gradient and the region with increased stress propagates in the opposite direction (Fig. 2b). The rupture slip velocity quickly reaches a maximum value of ~2 m/s, as expected for a regular EQ (Fig. 2c). Although slip velocity is predominantly in the direction of the global tangential stress, there is a small component in the $y$-direction as well (Fig. 2d). The maximum slip after ~4.1 s is about 5 m and the point of the maximum slip slitly shifts to the $+x$-direction (Fig. 2e). The rupture shape is asymmetric relative to $x = 0$ and is wider in the region $x > 0$.

Fig. 3 (curve 1) shows the seismic moment as a function of time for the rupture considered in Fig. 2. The dependence $M \propto T^3$ which develops after the short period of time needed for the slip velocity to reach a value of ~2 m/s is typical of regular EQs.

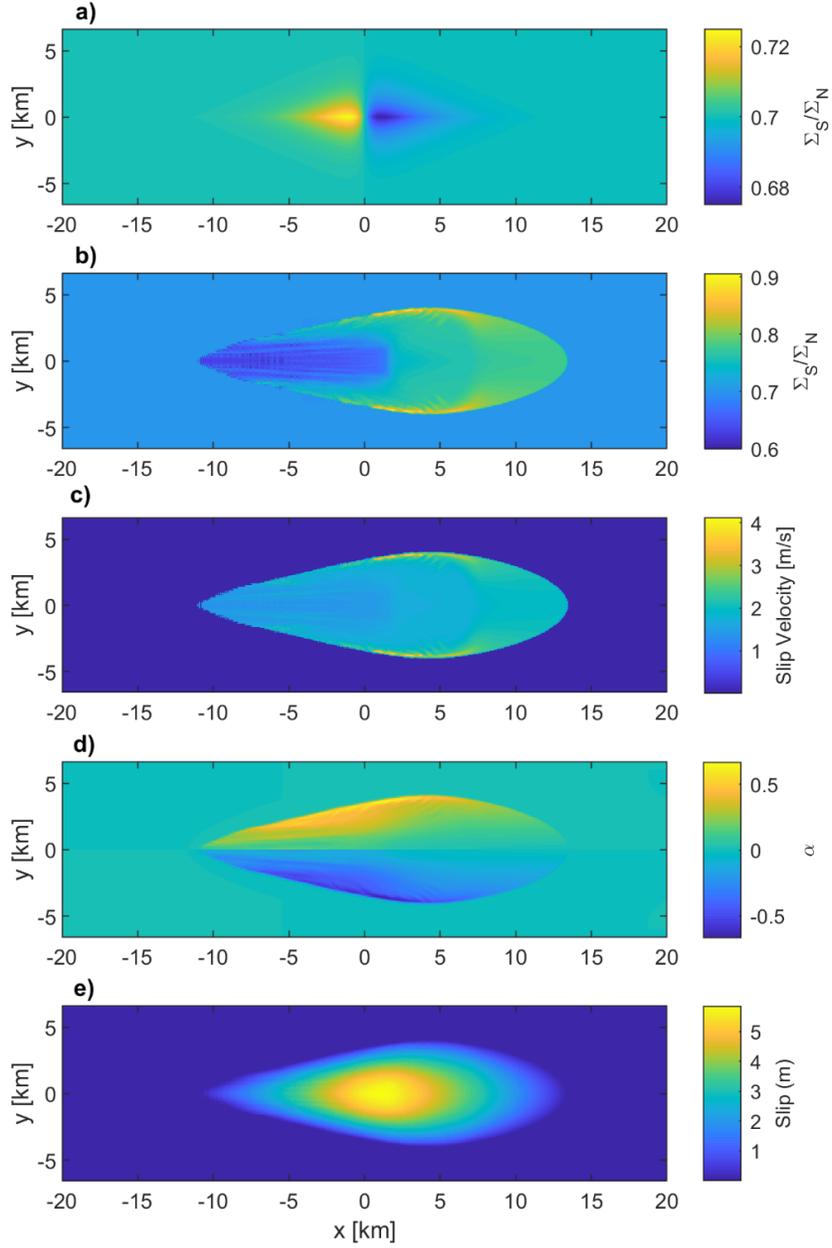

Figure 2. The initial stress inhomogeneity (a) is given by Eq. (16a) with parameters $\Sigma_{0S} \equiv \mu_0 = 0.7$, $\delta\Sigma_a = 0.1156$ (equivalent to $\delta\Sigma = 0.05$), $\zeta_x = 0.025$, $\zeta_{ya} = 0.05$, and $\beta = 0.1$. Model results, which are typical of a regular EQ, are shown for the spatial distribution of shear to normal stress ratio (b), slip velocity (c), angle between slip velocity and x-direction (d) and slip (e) at time 4.1 s after the beginning of a rupture. The global shear force is in the x-direction, imposing slip to the right.

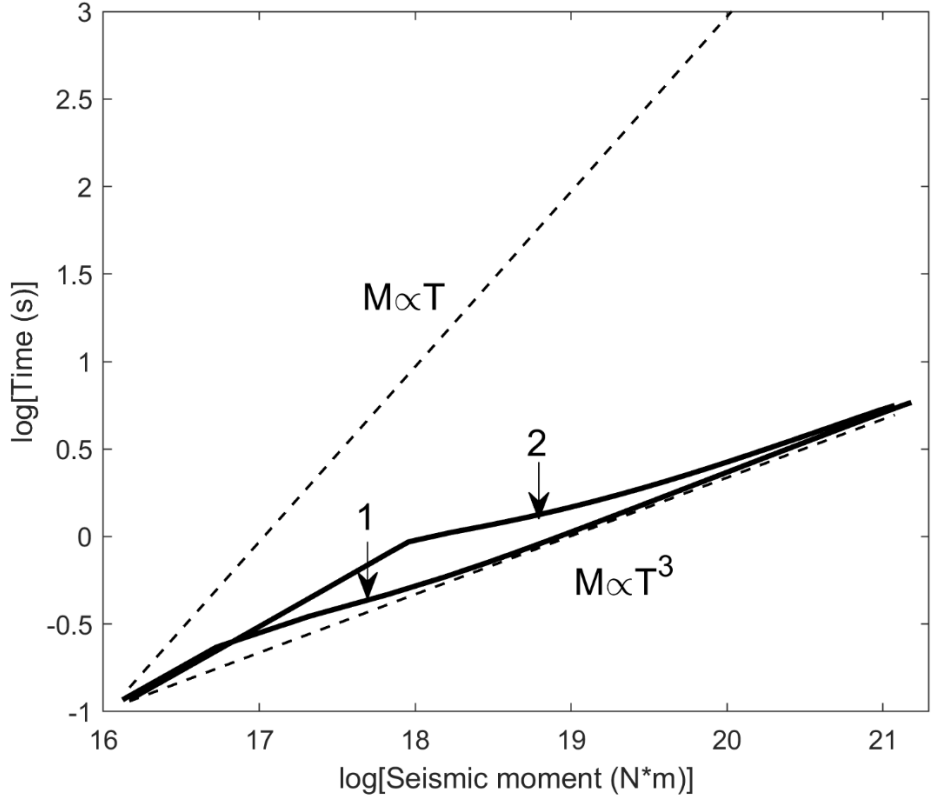

Figure 3. The seismic moment versus time dependence (solid lines) obtained for the rupture resulting from the initial shear stress shown in Fig. 2a (curve 1) and Fig 4a (curve 2).

The "hump-like" inhomogeneity in the initial stress given by Eq. 16$b$ (see also Fig. 4$a$), with $\Sigma_{0S} = 0.7$, $\delta\Sigma_b = 0.05$, $\zeta_x = 0.025$, and $\zeta_{yb} = 0.0207$, also demonstrates behavior typical of a regular EQ (Fig. 4). Again, we find 1) the rupture is a crack-like; 2) the region with reduced stress propagates in the direction of the initials stress gradient and the region with increased stress propagates in the opposite direction (Fig. 4b); 3) the maximum slip velocity in the rupture is about 2 m/s (Fig. 4c); 4) The maximum slip after ~4.1 s is about 5 m (Fig. 4d); 5) The crack is expanding in the $x$- and $y$-directions; 6) the seismic moment versus time dependence is $M \propto T^3$ (Fig. 3, curve 2). The only differences compared to the previous example with initial stress given by Eq. (16a) are 1) the rupture propagation velocity in the positive $x$-direction is visibly larger than in the negative $x$-direction and, respectively, 2) the shift of the point of the maximum slip to the +$x$-direction is larger.

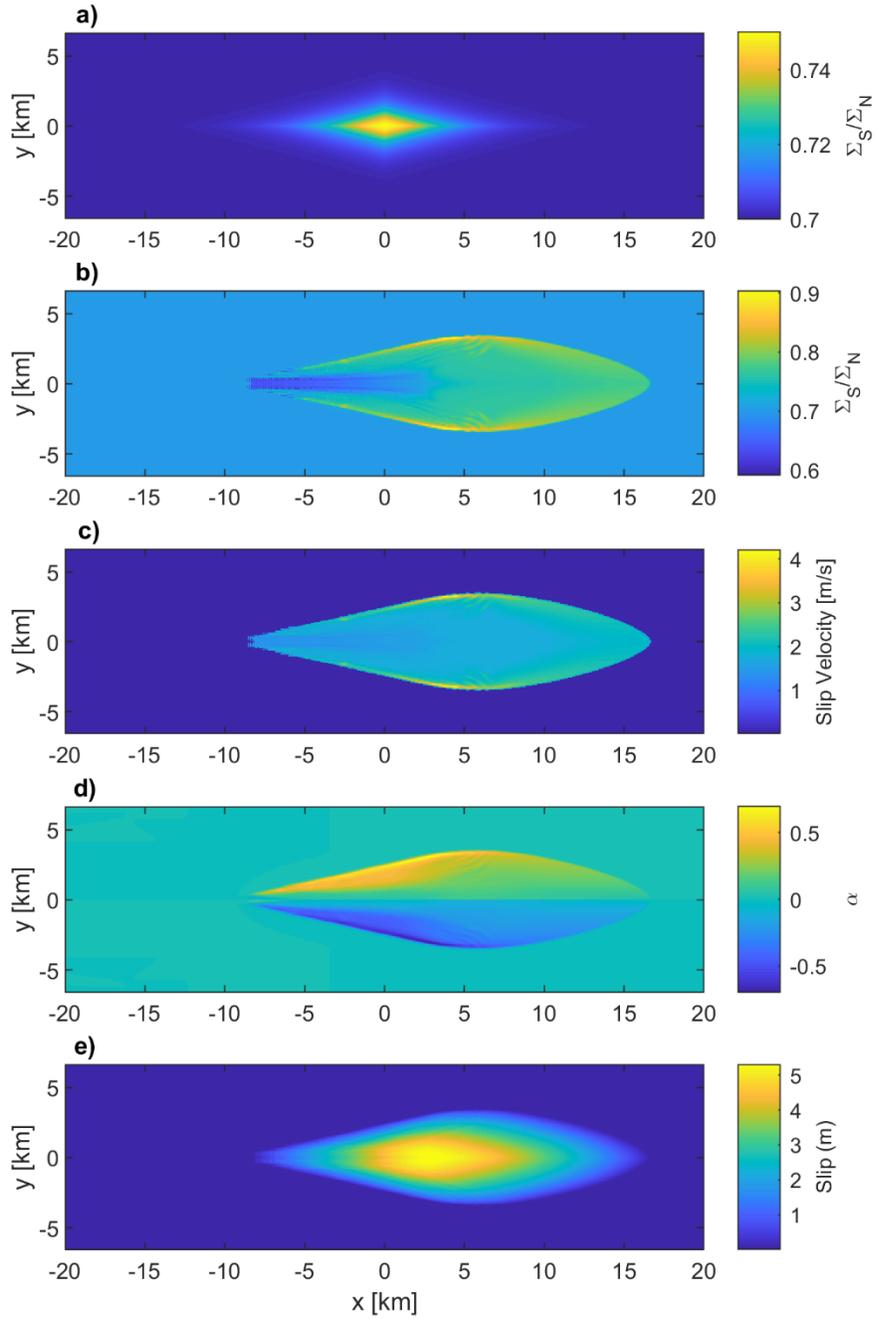

Figure 4. As in Fig. 2, but with the initial stress inhomogeneity (a) given by Eq. (16b) with parameters $\Sigma_{0S} \equiv \mu_0 = 0.7$, $\delta\Sigma_b \equiv \delta\Sigma = 0.05$, $\zeta_x = 0.025$, and $\zeta_{yb} = 0.0207$.

*SSEs*

Parameters of a seismic event crucially depend on the shear stress and stress gradient (see Fig. 6 in Part I). For the same shear stress as in the cases considered above but with a sufficiently small gradient in the stress, the rupture never reaches the slip velocity of 0.1 m/s needed for the abrupt reduction of friction (due to flash heating in our model, but other mechanisms are possible). The reduced friction required to increase the slip velocity to ~1 m/s, typical of EQs, doesn't occur and the rupture instead manifests as an SSE.

Figure 5 illustrates an example of an SSE which results from the initial stress profile of Eq. (16a), with parameters $\Sigma_{0S} = 0.7$ and a small gradient $\delta\Sigma_a = 0.01156$ (equivalent to $\delta\Sigma = 0.005$). The rupture is now pulse-like and localized in both the *x*- and *y*-directions, with a maximum slip velocity of ~1.2 cm/s. The rupture propagation velocity is ~1.1 km/s. Most significantly, note that there is no expansion of the pulse (rupture) in the *y*-direction. Therefore, the seismic moment versus time dependence is very different than for EQs (see Fig. 6). After a few seconds, which is the time needed to form two separate pulses, the seismic moment is directly proportional to time, $M \propto T$, which is the typical dependence for SSEs.

Figure 7 shows the results of simulating an SSE which is produced using the initial stress distribution of Eq. (16b) with parameters $\Sigma_{0S} = 0.2$ and, again, a small gradient resulting from $\delta\Sigma_a = 0.001$. As in the previous example, the rupture is also pulse-like and localized. The maximum slip velocity is ~2.6 cm/day, which is much smaller than in case above. The rupture propagation velocity is ~5 km/day. The seismic moment versus time dependence also approaches $M \propto T$ (Fig. 6) for the same reason, i.e., the size of the pulse remains the same during propagation.

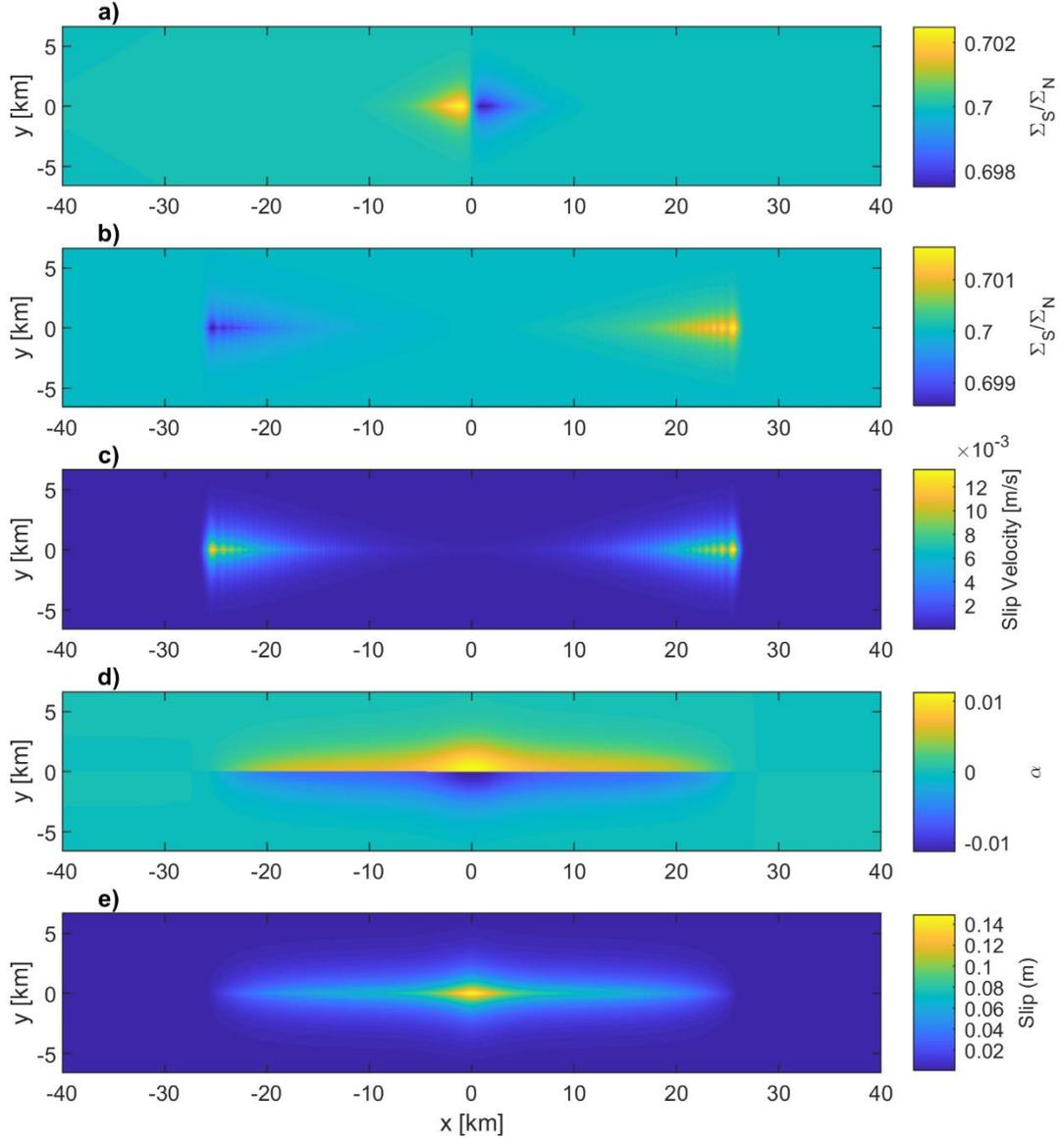

Figure 5. Similar to Fig. 2, i.e., with the initial stress inhomogeneity (a) given by Eq. (16a), but with parameters $\Sigma_{0S} \equiv \mu_0 = 0.7$, $\delta\Sigma_a = 0.01156$ (equivalent to $\delta\Sigma = 0.005$), $\zeta_x = 0.025$, $\zeta_{ya} = 0.05$, and $\beta = 0.1$. The gradient in the initial stress is now sufficiently small that the slip velocity remains below the threshold for flash heating to occur. Model results then give a SSE, as illustrated for the spatial distribution of shear to normal stress ratio (b), slip velocity (c), angle between slip velocity and x-direction (d) and slip (e) at time 23.3 s after the beginning of a rupture.

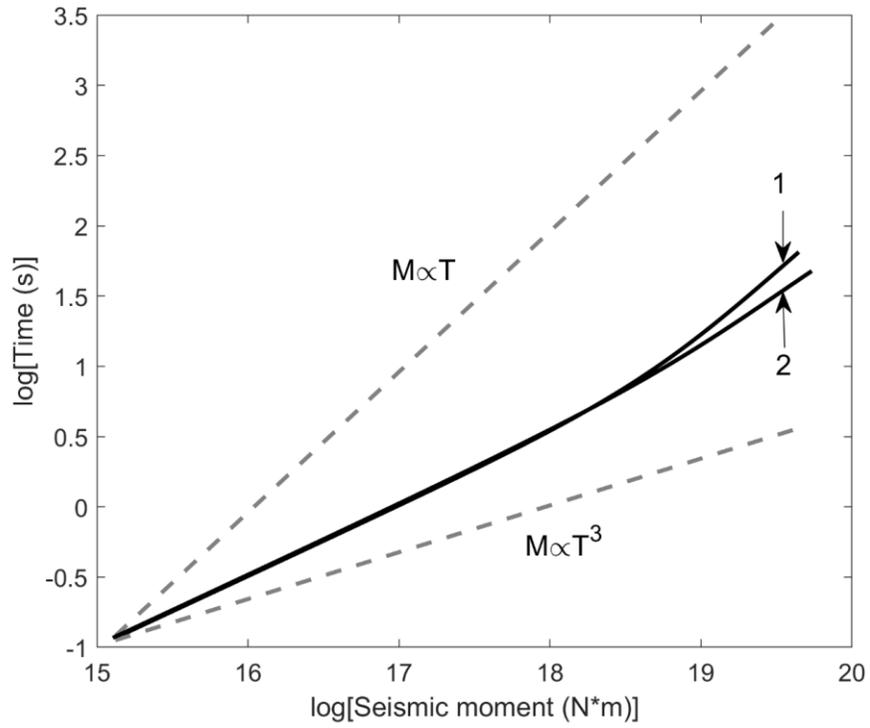

Figure 6. The seismic moment versus time dependence (solid lines) obtained for the rupture resulting from the initial shear stress shown in Fig. 5a (curve 1) and Fig 7a (curve 2).

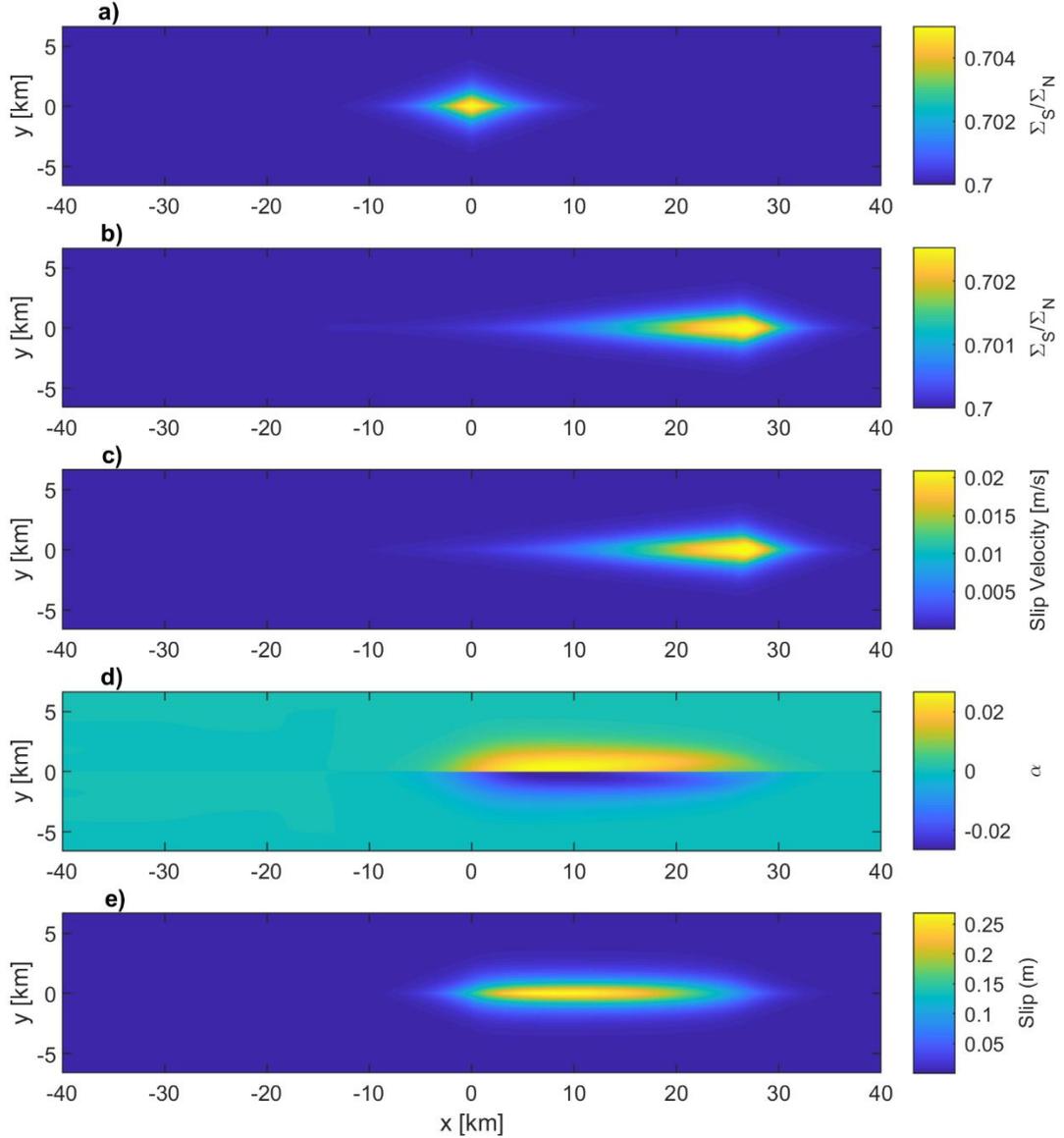

Figure 7. As in Fig. 5, but with the initial stress inhomogeneity (a) given by Eq. (16b) with parameters $\Sigma_{0S} = 0.7$, $\delta\Sigma_b = 0.005$, $\zeta_x = 0.025$, and $\zeta_{yb} = 0.0207$ again giving a SSE.

*Episodic Tremor and Slip*

Propagation of a slip pulse along a subduction fault is accompanied by massive bursts of non-volcanic tremor (NVT). This periodic phenomenon, known as episodic tremor and slip (ETS), has been observed virtually at all major subduction zones (Obara, 2002; 2009; Rogers, and Dragert, 2003; Kostoglodov et al., 2003; Kao et al., 2005; Rubinstein et al., 2010; Peng and Gomberg, 2010;

Gonzalez-Huizar et al., 2012). The tremor migration pattern during an ETS event is rather complicated (Shelly et al., 2007a, Obara, 2009; Ghosh et al., 2010(a); 2010(b); Houston et al, 2011). We have previously showed that our 1D model is able to account for some aspects of this pattern (Gershenzon et al, 2011; Gershenzon and Bambakidis, 2015). Here we continue the description of ETS phenomena within the context of the 2D model.

The commonly assumed scenario for ETS is that a slip pulse, generated at the boundary of a subduction fault at deep depth, propagates along the fault to a shallow depth with velocity ~10 km/day. An ETS event can be revealed by direct measurement via the GPS system and by monitoring accompanying bursts of NVT (Rogers, and Dragert, 2003). It is assumed that the stress disturbance accompanying a pulse produces tremors due to failure of small heterogeneities (asperities). This supposition is consistent with the observation that even small stress disturbances produced by seismic waves (from either the local medium or from distant large earthquakes or by tidal waves) may generate tremors (Miyazawa and Mori, 2006; Rubinstein et al., 2007, 2009; Peng et al.,2009; Miyazawa and Brodsky, 2008; Fry et al, 2011; Zigone et al., 2012; Chao et al, 2013; Rubinstein et al, 2008; Nakata et al, 2008; Thomas et al, 2009; Lambert et al, 2009). There are observations showing that sometimes the slip pulse produces tremors which migrate in the direction opposite to the direction of pulse propagation (Houston et al, 2011). This is the phenomenon of reverse tremor migration.

As shown in Fig. 7, stress heterogeneity with appropriate values of the effective shear stress and its gradient may produce a slip pulse with parameters that are typical of an ETS pulse. Since the width of a pulse in the propagation direction is usually much smaller than the size of the pulse front, we may model the initial condition given by formula (16b) with the same parameters as in Fig. 7 but with $\zeta_{yb} = 0$. As we know, after some time the pulse is formed and propagates in the positive *x*-direction. Let us suppose that there is an area with friction much larger than in the surrounding fault. Thus, the pulse could not rupture this particular area. Fig. 8 shows the result of simulations with such an obstacle. When the pulse reaches the "no rupture" zone, the stress wave is partially reflected producing reverse waves. The latter triggers tremors with reverse migration as observed (Houston et al, 2011).

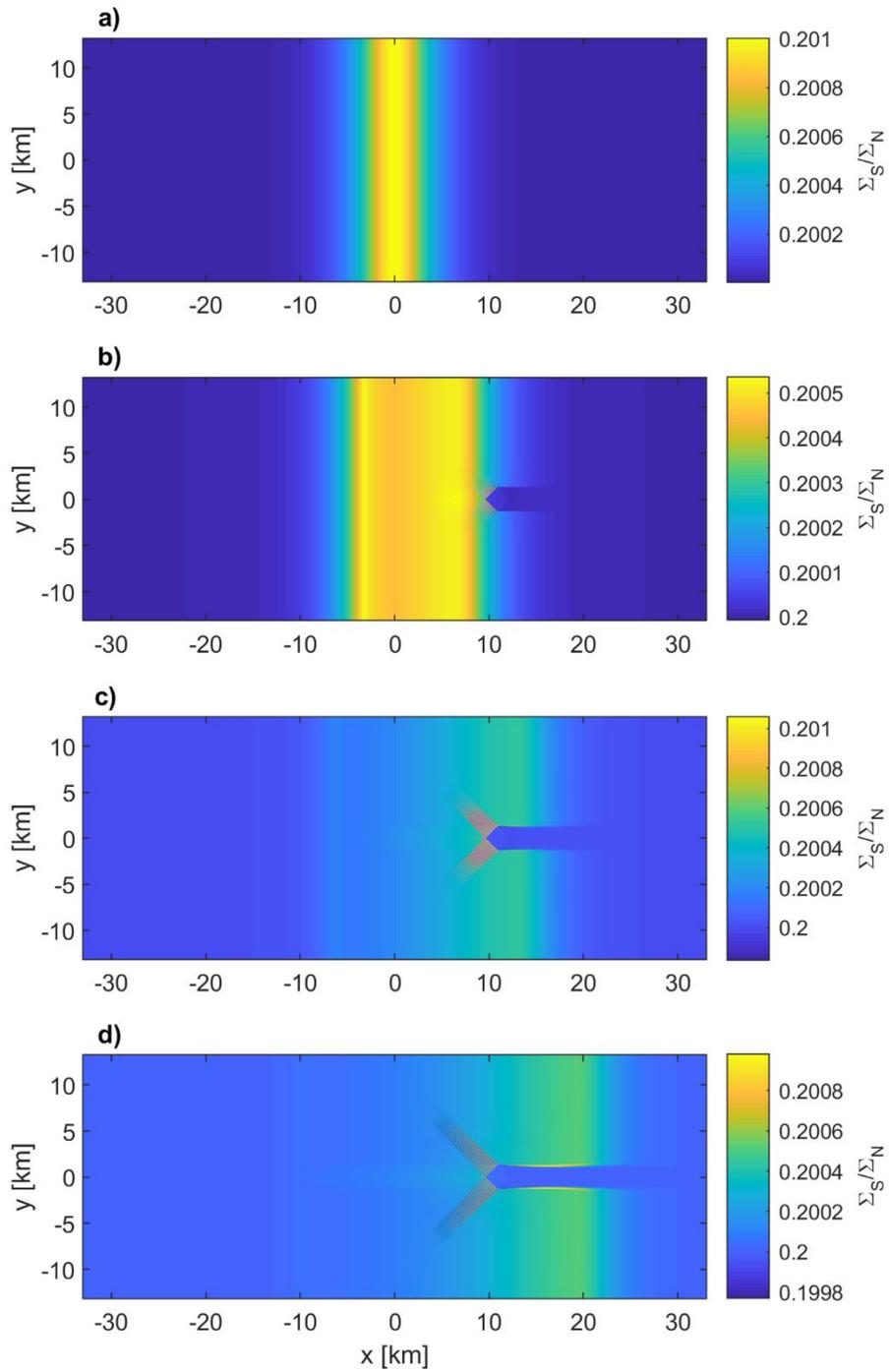

Figure 8. The spatial distribution of the shear to normal stress ratio is shown for different moments of time in the case where there is a localized region or obstacle with increased friction, located at $x \approx 11$ km, $y = 0$ km with a radius of ~1.3 km, that prevents a rupture from occurring in this region. The initial stress heterogeneity (a) is given by Eq. (16b) with parameters $\Sigma_{0S} = 0.2$, $\delta\Sigma_b = 0.001$, $\zeta_x = 0.025$, and $\zeta_{yb} = 0$. The model shows that the stress wave is partially reflected at the obstacle, resulting in a stress wave that can trigger tremors with the reverse migration which sometimes occurs in ETS events. (**a**) *t*=0, (**b**) *t*=1.4 days, (**c**) *t*=2.7 days, (**d**) *t*=4.1 days

**CONCLUSION**

The results of 2D simulations confirm the basic findings obtained by the 1D model, i.e., the type of a seismic event, EQ or SSE, and type of a rupture, crack-like or pulse-like, are determined by the fault strength, the ratio of the shear to normal stress, and the gradient in this ratio. The shape of a rupture and position of the maximum slip is determined by the initial stress spatial distribution. In addition to faithfully reproducing the 1D results, the 2D results show that the "head" of the rupture area in EQs is wider than the "tail" (Figs. 2 and 4). The slip velocity is primarily in the direction of the global shear stress direction, but it also has a component in the perpendicular direction (Fig 2e). The rupture expands in both the *x*- and *y*-directions in case of EQs (Figs. 2 and 4). In SSEs the rupture also extends in both directions for a short period of time, then, after pulses are formed, the shape of the pulse is practically unchanged (Figs. 6 and 8).

In contrast to the 1D model, the 2D model allows us to study the seismic moment versus time scaling law. Simulations show the scaling law may be expressed by the relation $M \propto T^\lambda$, where $1 < \lambda < 3$. For EQs, $\lambda = 3$ (see Figs. 3 and 5), while for SSEs, $\lambda$ tends to 1 after the formation of well-developed pulses (see Figs. 7 and 9). Such differences in scaling laws can be explained as follows. The seismic moment is expressed as the product of the characteristic rupture area *S* and the average slip *D*. For regular earthquakes, the rupture area expands in the *x* and *y*-directions at all times, so *D* and *S* are proportional to *T* and $T^2$, respectively, which explains the $M \propto T^3$ scaling law. For slow events, however, the pulse area does not change after the formation of a well-developed pulse, so the area remains constant in time, and we are only left with *D* proportional to *T*. Note, that at the beginning of EQ or SSE development, the value of $\lambda$ is neither 1 nor 3, suggesting that most generally, at arbitrary times, $1 < \lambda < 3$. This is confirmed by observations (see Fig.1 in Gomberg et al, 2016).

Finally, the 2D model provides not only a quantitative description for the parameters of a seismic pulse in ETS phenomena but also explains some features of non-volcanic tremor associated with this pulse, such as reverse tremor migration (Fig. 8).

**DATA AND RESOURCES**

All data used in this article came from published sources listed in the references.